\documentclass[aps,pra,floatfix,superscriptaddress,twocolumn,showpacs,10pt]{revtex4-1}
\usepackage{amssymb, amsmath,color,mciteplus,graphicx,subfigure}
\usepackage{calc,epsfig,epstopdf,color,mciteplus,bm,mathrsfs}
\usepackage{times}
\usepackage{float}
\usepackage{lipsum}

\begin{document}

\title{Noncyclic Geometric Quantum Gates with Smooth Paths via Invariant-based Shortcuts}
\author{Li-Na Ji}
\author{Cheng-Yun Ding}
\author{Tao Chen}
\affiliation{Guangdong Provincial Key Laboratory of Quantum Engineering and Quantum Materials, and School of Physics\\ and Telecommunication Engineering,  South China Normal University, Guangzhou 510006, China}

\author{Zheng-Yuan Xue} \email{zyxue83@163.com}
\affiliation{Guangdong Provincial Key Laboratory of Quantum Engineering and Quantum Materials, and School of Physics\\ and Telecommunication Engineering,  South China Normal University, Guangzhou 510006, China}

\affiliation{Guangdong-Hong Kong Joint Laboratory of Quantum Matter, and Frontier Research Institute for Physics,\\ South China Normal University, Guangzhou  510006, China}

\begin{abstract}
 Nonadiabatic geometric quantum computation is dedicated to the realization of high-fidelity and robust quantum gates, which are necessary for fault-tolerant quantum computation. However, it is limited by cyclic and mutative evolution path, which usually requires longer gate-time and abrupt pulse control, weakening the gate performance. Here, we propose a scheme to realize geometric quantum gates with noncyclic and nonadiabatic evolution via invariant-based shortcuts, where universal quantum gates can be induced in one step without path mutation and the gate time is also effectively shortened. Our numerical simulations show that, comparing with the conventional dynamical gates, the constructed geometric gates have stronger resistance not only to systematic errors, induced by  both qubit-frequency drift and the deviation of the amplitude  of the driving fields,  but also to  environment-induced decoherence effect. In addition, our scheme can also be implemented on a superconducting circuit platform, with the fidelities of single-qubit and two-qubit gates are higher than 99.97$\%$ and 99.84$\%$, respectively. Therefore, our scheme provides a promising way to realize high-fidelity fault-tolerant quantum gates for scalable quantum computation.
\end{abstract}
\maketitle

\section{Introduction}
Quantum computers can efficiently handle some hard problems\cite{QC001, QC002} that the classical ones can not, due to its intrinsic massive quantum parallel computation nature. The basic necessity for quantum computation is to implement a set of universal quantum gates based on quantum mechanical principles\cite{QC002}. However, the coherence of quantum systems cannot be maintained perfectly, due to the inevitable interaction with their surrounding environment, which leads to the fact that any quantum gate can not be perfect and the operation time for a quantum task is limited. Therefore, to realize large-scale fault-tolerant quantum computation, it is challenging and significant to complete quantum gates with high fidelity  and strong robustness, under the limited coherent times.

It is well-known that the adiabatic process finds many applications in modern physics\cite{add1, add2, add3, add4, add5}. Remarkably,  in 1984, geometric phase under adiabatic and cyclic evolution conditions was discovered by Berry\cite{Berry003}, which only depends on the global feature of the evolution path and is insensitive to the detail of the accompanied noises. So, quantum computation with quantum gates that are constructed by geometric phases, the so-called geometric quantum computation\cite{A004, A005}, has the distinct merit of being resilient to certain noises. Subsequently, elementary geometric quantum gates based on adiabatic Abelian\cite{Berry003} and non-Abelian\cite{A009} geometric phases have been demonstrated \cite{ExperA010, ExperA011, ExperA012, ExperA013}. However, this slow adiabatic process may leads to serious gate infidelity, due to the prolonged exposure of the target quantum system into the environment.

In 1987, Aharonov and Anandan\cite{AAP013} proposed to induce geometric phase without the adiabatic condition. Then the nonadiabatic geometric quantum computation (NGQC) schemes based on Abelian\cite{NG015, NG016, NG017} and non-Abelian\cite{NH020, Xugf} geometric phases have been proposed. NGQC uses the much faster nonadiabatic process, and thus it has soon attracted much attention, and up to now, it has been experimentally implemented in various systems, such as trapped ions\cite{trapped029, trapped2, trapped3}, superconducting quantum circuits\cite{super025, super026, super027, super028},  nuclear magnetic resonance\cite{NMR032, NMR033, NMR2, NMR034} and nitrogen-vacancy centres\cite{NV033, NV2, NV3, NV4, NV5}, etc.
Currently, the orange-slice-shaped geometric path (OSSP)\cite{Orange043} is widely used in NGQC\cite{NG019, NG020, Orange044, Orange045, NH024} for single-shot realization of arbitrary quantum gates, which undergoes a multi-segment process with different parameters. However, comparing with the dynamical quantum gates from Rabi oscillation, this implementation has the following drawbacks. Firstly, due to the cyclic condition, the needed gate-time is at least twice of that of the dynamical ones. Besides, the gate-time is the same for any possible gate, even a very small rotation. Secondly, it requires abrupt change of the Hamiltonian parameters, that determine the evolution path, which increase the gate infidelity and the complexity of experimental control.

To avoid these drawbacks, the noncyclic geometric phase\cite{SB034} could help. It not only shortens the evolution path, i.e., shortens the gate-time, but also brings flexibility to the evolution path design in constructing quantum gates\cite{NNGQC039, NNGQC040, NNGQC041, duyx, NNGQC042}. Here, we propose a scheme for noncyclic NGQC with smooth geometric path and detail its implementation on superconducting quantum circuits. In our scheme, we use invariant-based inverse engineering\cite{LR046, LR047, LR048} to construct the governable Hamiltonian, aiming at finding shorter noncyclic geometric path for arbitrary rotation operations, and thus the decoherence-induced gate infidelity can be effectively  suppressed. In addition, we succeed in avoiding the abrupt pulse control in constructing the geometric path, which can decrease the error caused by parameter mutation and the demands of experimental realization. Different from the unconventional geometric phase scheme in Ref. \cite{NNGQC042}, we here completely cancel the accompanied dynamical phases, thus obtain quantum gates based on the pure geometric phase. Besides, we can achieve arbitrary X-, Y- and Z-axis rotation operations in one step. Due to these merits, the numerical simulations show that our scheme can improve the gate performance remarkably compared with the cyclic NGQC and dynamical schemes. Finally, we implement our scheme on a superconducting circuit consists of transmon qubits, where the gate fidelities of single-qubit gates and two-qubit gate can be as high as 99.97$\%$ and 99.84$\%$, respectively. Therefore, our scheme provides a promising way to fault-tolerant quantum computation.

\section{Noncyclic geometric quantum gates}
In this section, we present our protocol for implementing geometric quantum gates with noncyclic nonadiabatic geometric phases. In our scheme, the smooth and shorter trajectories of  geometric phases are founded. We final evaluate our gate performance in terms of gate fidelity and robustness.

\subsection{The geometric phase}

\begin{figure}
\includegraphics[width=8.5cm]{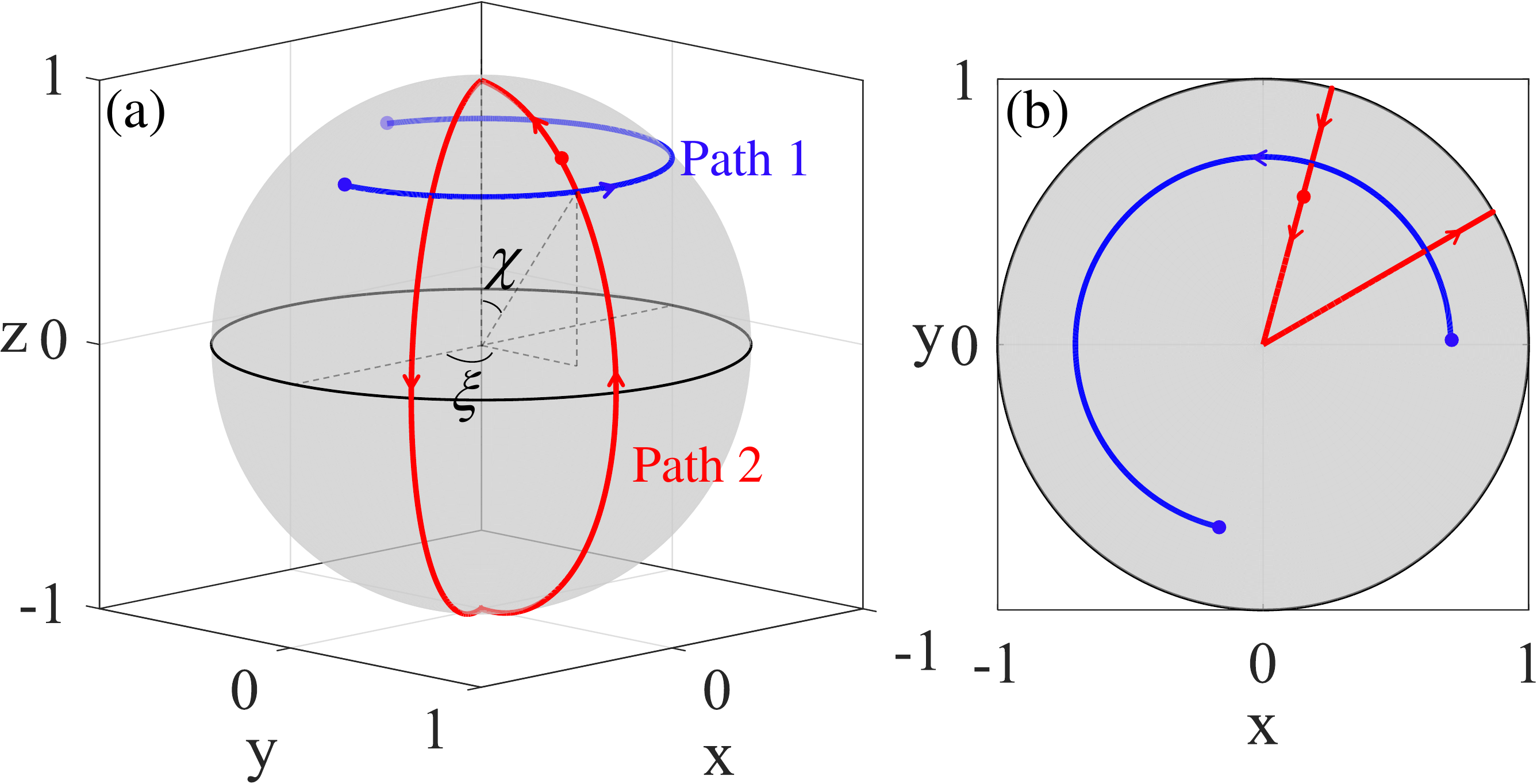}
\caption{The visualized evolution trajectories of state $|\psi_{1}(t)\rangle$ on a Bloch sphere. $\chi$ and $\xi$ represent the polar angle and azimuth angle of evolution states, respectively. The two blue dots represent initial and final positions . (a) The trajectories of our NSGP (Path 1) and cyclic OSSP (Path 2) schemes.  (b) Top view of diagram (a), Path 2 has mutation points obviously, while Path 1 is smooth.}\label{fig1}
\end{figure}

Consider a two-level system with lower and upper energy levels denoted by $\{|0\rangle=(1,\,0)^\dag, |1\rangle=(0,\,1)^\dag\}$, with their frequency difference being $\omega_0$, their transition   is induced by a microwave field with coupling strength $\Omega(t)$, frequency $\omega_d(t)$ and phase $\phi(t)$.  Assume $\hbar=1$ hereafter, in the rotating framework, the interaction Hamiltonian  is
\begin{equation}
\label{01}
\begin{split}
\mathcal{H}(t)
=&\frac{1}{2}\left(
  \begin{array}{cc}
    -\Delta(t)& \Omega(t)e^{-i\phi(t)} \\
    \Omega(t)e^{i\phi(t)} & \Delta(t) \\
  \end{array}
\right),
\end{split}
\end{equation}
where $\Delta(t)=\omega_0-\omega_d(t)$ is a detuning, which can be time-dependent in general. To implement the nonadiabatic process under the Hamiltonian $\mathcal{H}(t)$, we then take a pair of orthogonal dressed states
\begin{equation}
\label{02}
\begin{split}
|\psi_{1}(t)\rangle &=\cos\frac{\chi(t)}{2}|0\rangle+\sin\frac{\chi(t)}{2}e^{i\xi(t)}|1\rangle,\\
|\psi_{2}(t)\rangle &=\sin\frac{\chi(t)}{2}e^{-i\xi(t)}|0\rangle-\cos\frac{\chi(t)}{2}|1\rangle
\end{split}
\end{equation}
as a set of evolution states, which are the eigenstates of the Lewis-Riesenfeld invariant of\cite{LR046, LR047, LR048}
\begin{equation}
\label{03}
I(t)=\frac{\mu}{2}\left(
                    \begin{array}{cc}
                      \cos\chi(t) & \sin\chi(t)e^{-i\xi(t)} \\
                      \sin\chi(t)e^{i\xi(t)} & -\cos\chi(t) \\
                    \end{array}
                  \right),
\end{equation}
with $\mu$ being an arbitrary constant. By solving dynamic equation $i\partial I(t)/\partial t-[\mathcal{H}(t),I(t)]=0$, the parameter relationships between $\{\Omega(t),\Delta(t),\phi(t)\}$ describing $\mathcal{H}(t)$ and $\{\chi(t),\xi(t)\}$ describing ${|\psi_{1,2}(t)\rangle}$ can be obtained as
\begin{equation}
\begin{split}
\label{04}
&\dot\chi(t)=\Omega(t)\sin[\phi(t)-\xi(t)],\\
&\dot\xi(t)=-\Delta(t)-\Omega(t)\cot\chi(t)\cos[\phi(t)-\xi(t)],
\end{split}
\end{equation}
where parameters $\chi(t)$ and $\xi(t)$ in Equation (\ref{02}) can be regarded as the polar and azimuth angles on a Bloch sphere with $\chi(t)\in[0,\pi]$ and $\xi(t)\in[0,2\pi)\pm2n\pi$ with $n=0,1,2,...$. Thus the evolution trajectories   of the  states $|\psi_{1,2}(t)\rangle$ can be visualized on a Bloch sphere, as shown in Figure \ref{fig1}. Therefore, the Hamiltonian $\mathcal{H}(t)$ can be inversely engineered, according to the restricted requirements of parameters $\chi(t)$ and $\xi(t)$ under our targeted evolution path. In this way, we next aim to achieve the noncyclic and nonadiabatic geometric evolution path based on the above invariant-based shortcuts.

Furthermore, by setting the states $|\psi_{1}(t)\rangle=e^{-i\gamma(t)}|\Psi_{1}(t)\rangle$ and $|\psi_{2}(t)\rangle=e^{i\gamma(t)}|\Psi_{2}(t)\rangle$ with $\gamma(0)=0$, where $|\Psi_{1,2}(t)\rangle$ satisfy Schr\"{o}dinger equation $i\partial |\Psi_{1,2}(t)\rangle/\partial t\!=\!\mathcal{H}(t)|\Psi_{1,2}(t)\rangle$, thus after an  evolution time interval $\tau$, we can get the overall phase as
\begin{equation}\label{05}
\gamma(\tau)=\frac{1}{2}\int_0^\tau\frac{\Delta(t)+\dot\xi\sin^2\chi}{\cos\chi}dt
-\frac{1}{2}\int_0^\tau(1-\cos\chi)\dot\xi dt.
\end{equation}
And the corresponding  evolution operator is
\begin{equation}\label{06}
\begin{split}
U(\tau)=&|\Psi_{1}(\tau)\rangle \langle \Psi_{1}(0)|+|\Psi_{2}(\tau)\rangle \langle \Psi_{2}(0)|\\
=&e^{i\gamma(\tau)}|\psi_{1}(\tau)\rangle \langle \psi_{1}(0)|+e^{-i\gamma(\tau)}|\psi_{2}(\tau)\rangle \langle \psi_{2}(0)|\\
=&\left(
  \begin{array}{cc}
    u_1& u_2\\
   -u_2^*& u_1^*\\
  \end{array}\right),\\
\end{split}
\end{equation}
where
\begin{equation}\label{006}
\begin{split}
 u_1=&(\cos\Gamma\cos\frac{\chi_{-}}{2}+i\sin\Gamma\cos\frac{\chi_{+}}{2})e^{-i\frac{\xi_{-}}{2}},\\
 u_2=&(-\cos\Gamma\sin\frac{\chi_{-}}{2}+i\sin\Gamma\sin\frac{\chi_{+}}{2})e^{-i\frac{\xi_{+}}{2}},
\end{split}
\end{equation}
with $\xi_{\pm}=\xi(\tau)\pm\xi(0)$, $\chi_{\pm}=\chi(\tau)\pm\chi(0)$ and $\Gamma=\gamma(\tau)+\xi_{-}/2$.

For a general noncyclic evolution path, the dynamical part of the overall phase can be calculated as
\begin{equation}
\label{08}
\begin{split}
\gamma_d(\tau)&=-\int_0^\tau \langle \psi_{1}(t)|\mathcal{H}(t)|\psi_{1}(t)\rangle dt\\
&=\frac{1}{2}\int_0^\tau\frac{\Delta(t)+\dot\xi\sin^2\chi}{\cos\chi}dt.
\end{split}
\end{equation}
and the geometric phase of a noncyclic path $\mathcal{C}_1$ with its geodesic line $\mathcal{C}_2$ connecting the initial and final points of the actual evolution path is
\begin{equation}
\begin{split}
\label{11}
\gamma_g'=\gamma_g^{\mathcal{C}_1}+\gamma_g^{\mathcal{C}_2}
=-\frac{1}{2}\oint_{\mathcal{C}_1+\mathcal{C}_2}(1-\cos\chi)\dot\xi dt,
\end{split}
\end{equation}
which can be converted into a surface integral, see Appendix A for details. Therefore, the geometric property of $\gamma_g'$ can be explained by the solid angle bounded by $\mathcal{C}_1$ and $\mathcal{C}_2$.

It is generally believed that geometric phase has an intrinsic positive contribution to the resistance of local noise. Ref. \cite{NNGQC042} implements geometric quantum gates in the presence of nonzero dynamical phase, i.e.,  the unconventional geometric gates. In contrast, we here try to completely eliminate the dynamical phase, $\gamma_d=0$ in Equation (\ref{08}), and achieve arbitrary geometric operations based on pure geometric phase. For our purpose, one method is to let the dynamical phase equals zero all the time, i.e.,
\begin{equation}\label{Deltat}
\Delta(t)=-\dot\xi(t)\sin^2\chi(t),\,\,t\in[0, \tau].
\end{equation}
Another method is to null the dynamical phase at the final time, in this case, the detuning parameter $\Delta(t)$ can be chosen as a constant of
\begin{equation}\label{Delta}
\Delta(t)\equiv\Delta=-{1 \over \tau} \int_0^\tau\dot\xi(t)\sin^2\chi(t) dt.
\end{equation}
Remarkably, in the latter method, we do not need to tune the driving or qubit frequency in a time-dependent way, removing the need of deliberate control in the former case, and thus simplifies the  complex quantum  control.

\subsection{Smooth geometric gates}
After eliminating the dynamical phase, universal geometric quantum gates can be realized by assigning $\Gamma$, $\chi_{\pm}$ and $\xi_{\pm}$ in Equation (\ref{06}). However,  different choices of the slope of $\chi(t)$ and $\xi(t)$ correspond to different evolution paths for a same gate. In the following, we present a new scheme of NGQC using  noncyclic geometric phases, and the  path of which evolves along the latitude line on Bloch sphere. Besides, the path is smooth without the mutation of control parameters, and thus further simplifies the  complex quantum  control. Based on noncyclic smooth geometric path (NSGP) scheme we designed below, a set of arbitrary and equivalent single-qubit geometric X-, Y- and Z-axis rotation operations, denoted as $R_x(\theta_x)$-like, $R_y(\theta_y)$-like and $R_z(\theta_z)$ gates, can be realized, where $\theta_{x,y,z}$ are rotation angles.

Geometric property in Equation (\ref{11}) holds when $\dot\xi(t)\neq0$ and $\chi\neq0$; besides, $\chi\neq\pi/2$ when cancel the dynamical phase, see Equation (\ref{08}). Based on the above  conditions, we consider a NSGP along the latitude line, as Path 1 shown in Figure \ref{fig1}. When setting $\dot\chi=0$,  the corresponding evolution operator elements in Equation (\ref{006}) can be simplified to
\begin{equation}
\label{12}
\begin{split}
 u_1=&(\cos\Gamma+i\sin\Gamma\cos\chi)e^{-i\frac{\xi_{-}}{2}},\\
 u_2=& i\sin\Gamma\sin\chi e^{-i\frac{\xi_{+}}{2}},
\end{split}
\end{equation}
where $\Gamma=\xi_{-}\cos\chi/2$. And
\begin{equation}
\gamma(\tau)= {1 \over 2}  \int_0^\tau(\cos\chi -1)\dot\xi dt= {\xi_{-} \over 2} (\cos\chi-1)
\end{equation}
due to $\gamma_d=0$. In this way, $R_x(\theta_x)$-like, $R_y(\theta_y)$-like and $R_z(\theta_z)$ rotations  can be achieved by setting
\begin{equation}
\begin{split}
\Gamma&=\pi/2,\ \xi(0)=\pi/2,\ \chi=\theta_x/2;\\
\Gamma&=\pi/2,\ \xi(0)=\pi,\ \chi=\theta_y/2;\\
\Gamma&=\pi,\ \xi_{-}=\theta_z+2\pi,
\end{split}
\end{equation}
respectively. 
Note that, $R_{x,y}(\theta_{x,y})$-like gates include a $R_z$($\xi_{-}'$) operation, with $\xi_{-}'=\xi_{-}-\pi$, in front of the target $R_{x,y}(\theta_{x,y})$ rotations, which can also be used to construct universal set of quantum gates and can be cancelled out by an inverse operation $R_z(-\xi_{-}')$. We now detail the realization of a set of universal single-qubit gates in the following. Hadamard-like gate can be realized by setting $\Gamma=\pi/2, \xi(0)=0$ and $\chi=\pi/4$ including a $R_z(\sqrt2\pi)$ operation in front of itself. Phase gate and $\pi/8$ gate can be realized by $R_z(\pi/2)$ and $R_z(\pi/4)$, respectively.

\begin{figure}
\includegraphics[width=8.5cm]{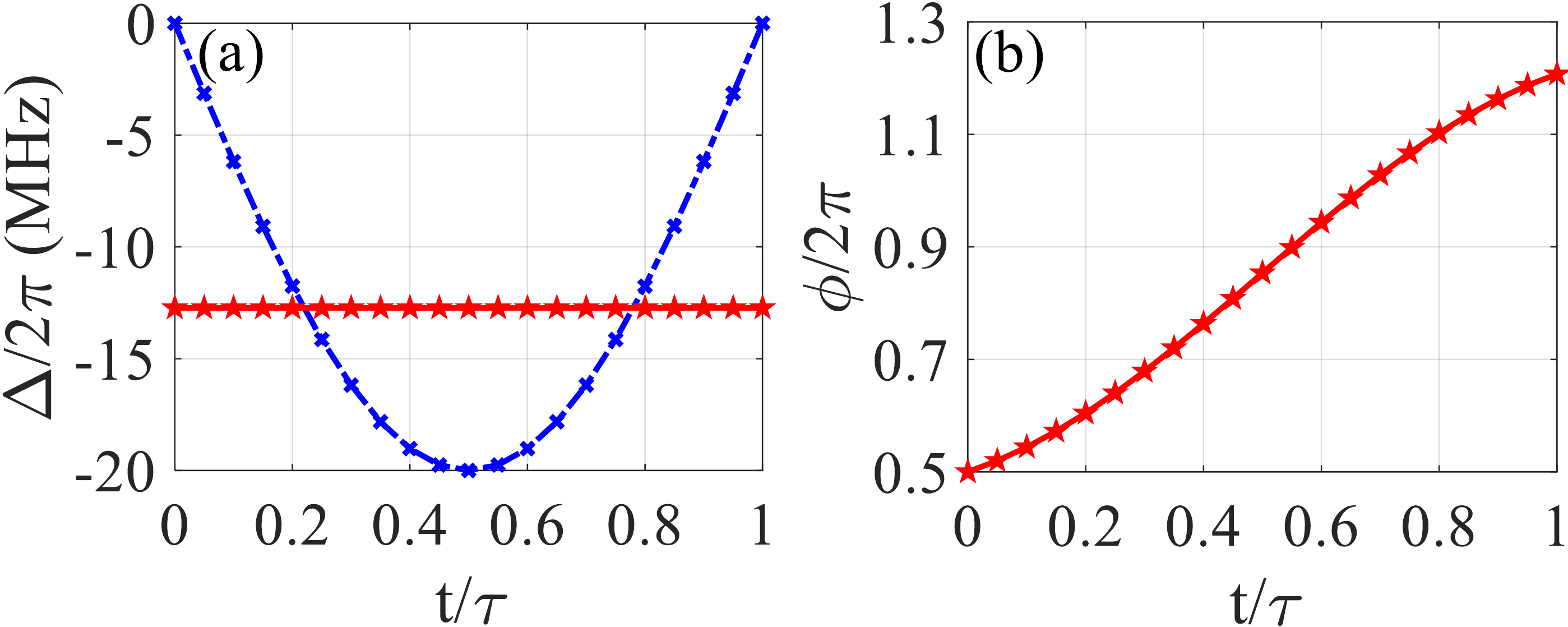}
\caption{The slopes of the Hamiltonian parameters for Hadamard-like gate, for the case of $\dot\chi=0, \phi(t)-\xi(t)=\pm\pi$. We set $\Omega(t)=\Omega_{\rm m}\sin(\pi t/\tau)$ with $\Omega_m=2\pi\times20$ MHz here. (a) The slopes of the detuning for eliminating the dynamical phase according to Equation (\ref{Deltat}) (the blue line) and Equation (\ref{Delta}) (the red line). (b) The slopes of the phase of the driving field.}
\label{fig2}
\end{figure}

In the following, based on the above setting of path parameters, we inversely engineer the Hamiltonian parameters $\Delta$ and $\phi(t)$ according to Equations (\ref{04}) and (\ref{Delta}). Defining $\phi(t)=\phi_0+\phi_1(t)$ with $\phi_0$ is a constant and $\phi_1(0)=0$, and choosing $\dot\chi=0$, so that $\phi(t)=\xi(t)\pm\pi$, the slopes of $\Delta$ and $\phi_1(t)$ are found to be
\begin{equation}\label{13}
\begin{split}
&\Delta=-{1 \over \tau }\int_0^{\tau}\Omega(t)\tan\chi dt,\\
&\phi_1(t)=-\Delta t +\int_0^t\Omega(t)\cot\chi dt,
\end{split}
\end{equation}
where the pulse shape $\Omega(t)$ can be selected according to certain experimental systems and/or optimization purpose. Setting $\Omega(t)=\Omega_{\rm m}\sin(\pi t/\tau)$, which is easy achieved experimentally, we plot the shapes of $\Delta$ and $\phi(t)$ of Hadamard-like gate  in Figure \ref{fig2}.

\subsection{Gate performance}

We further considered our gate performance under the decoherence effect, by numerical simulation of the  Lindblad master equation\cite{Master048} of
\begin{equation}\label{16}
\begin{split}
\dot\rho_1=-i[\mathcal{H}(t),\rho_1]+\kappa_1\mathcal{A}(b_{1-})+\kappa_2\mathcal{A}(b_{1z}),
\end{split}
\end{equation}
where $\rho_1$ is the density operator, $\kappa_1$ and $\kappa_2$ are represented as the decay and dephasing rates, the Lindblad operator $\mathcal{A}(b)=2b\rho_1 b^{\dagger}-b^{\dagger} b \rho_1-\rho_1 b^{\dagger} b$, $b_{m-}=\sum_{k=1}^{+\infty}\sqrt k |k-1\rangle_m\langle k|$ and $b_{mz}=\sum_{k=1}^{+\infty}k |k\rangle_m\langle k|$ are the standard lower operator and the projector of $k${\rm th} level for $m$th qubit, respectively. Consider an ideal two-level system generalized by single qubit here, then $k= m=1$. To fully test the performance of the implemented quantum gates, we set the initial state in  a general form of $\psi_i=\cos\theta|0\rangle+\sin\theta|1\rangle$. Accordingly, under Hadamard-like and $\pi/8$ gate operations, the final states $\psi_{\tau}$ are $[(\cos\theta+\sin\theta)|0\rangle+(\cos\theta-\sin\theta)\exp(i\sqrt2\pi)|1\rangle]/\sqrt2$ and $\cos\theta|0\rangle\!+\!\sin\theta\exp(i\pi/4)|1\rangle$, respectively. We define gate fidelity as $F_G=\int_0^{2\pi}\langle \psi_{\tau}|\rho_1|\psi_{\tau}\rangle d\theta/(2\pi)$\cite{fidelity} with the integration being numerically done for 1001 initial states where $\theta$ is evenly distributed over [0, 2$\pi$]. In the following, to prove the superiority of our scheme, we consider the cyclic geometric gates from the OSSP (see Appendix B) and the dynamical path (DP) schemes (see Appendix C) as references. Figure \ref{fig4} shows the reduction of gate fidelity due to decoherence for these three schemes, where $\kappa=\kappa_1=\kappa_2$. One can see that the line for our scheme is very flat; besides when $\kappa= 10^{-3}\Omega_\textrm{m}$, the gate fidelity is still higher than 99.90$\%$. Meanwhile, we also calculate the effective pulse area $S=\int_0^{\tau}\Omega(t)dt/2$ for our scheme, and $S=\sqrt2\pi/4, 3\pi/5$ and $\sqrt{17}\pi/9$ for Hadamard-like gate, Phase gate and $\pi/8$ gate, respectively. Instead, for dynamical scheme, they are $3\pi/4, 3\pi/4$ and $5\pi/8$, and for OSSP scheme, they are all equals $\pi$. Thus, we conclude  that our scheme has the shortest gate-time, which is   the most important factor  contributing to our excellent gate performance.

\begin{figure}
\includegraphics[width=8.5cm]{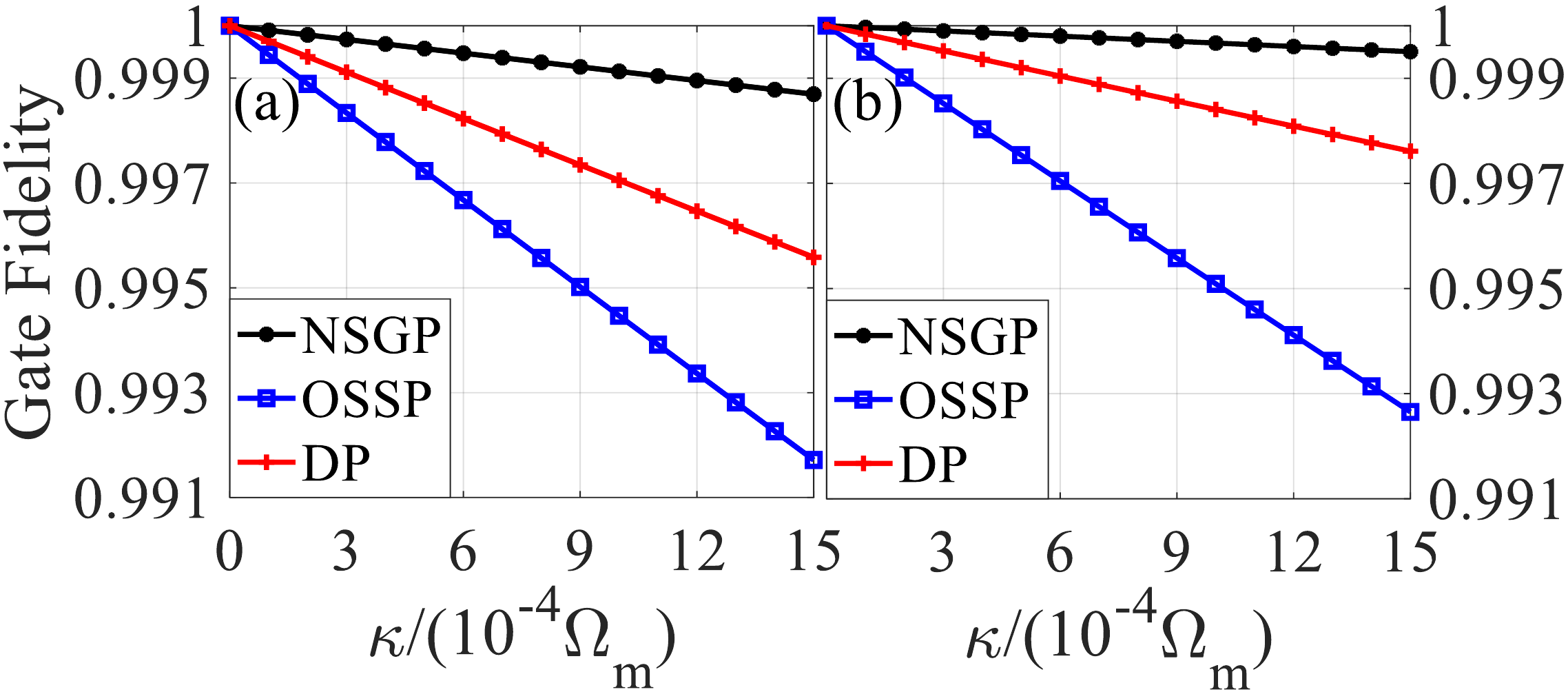}
\caption{Gate fidelities for (a) Hadamard-like and (b) $\pi/8$ gates as function of decoherence rate $\kappa$ for the current NSGP, OSSP and DP schemes.} \label{fig4}
\end{figure}

\begin{figure}
\includegraphics[width=8.5cm]{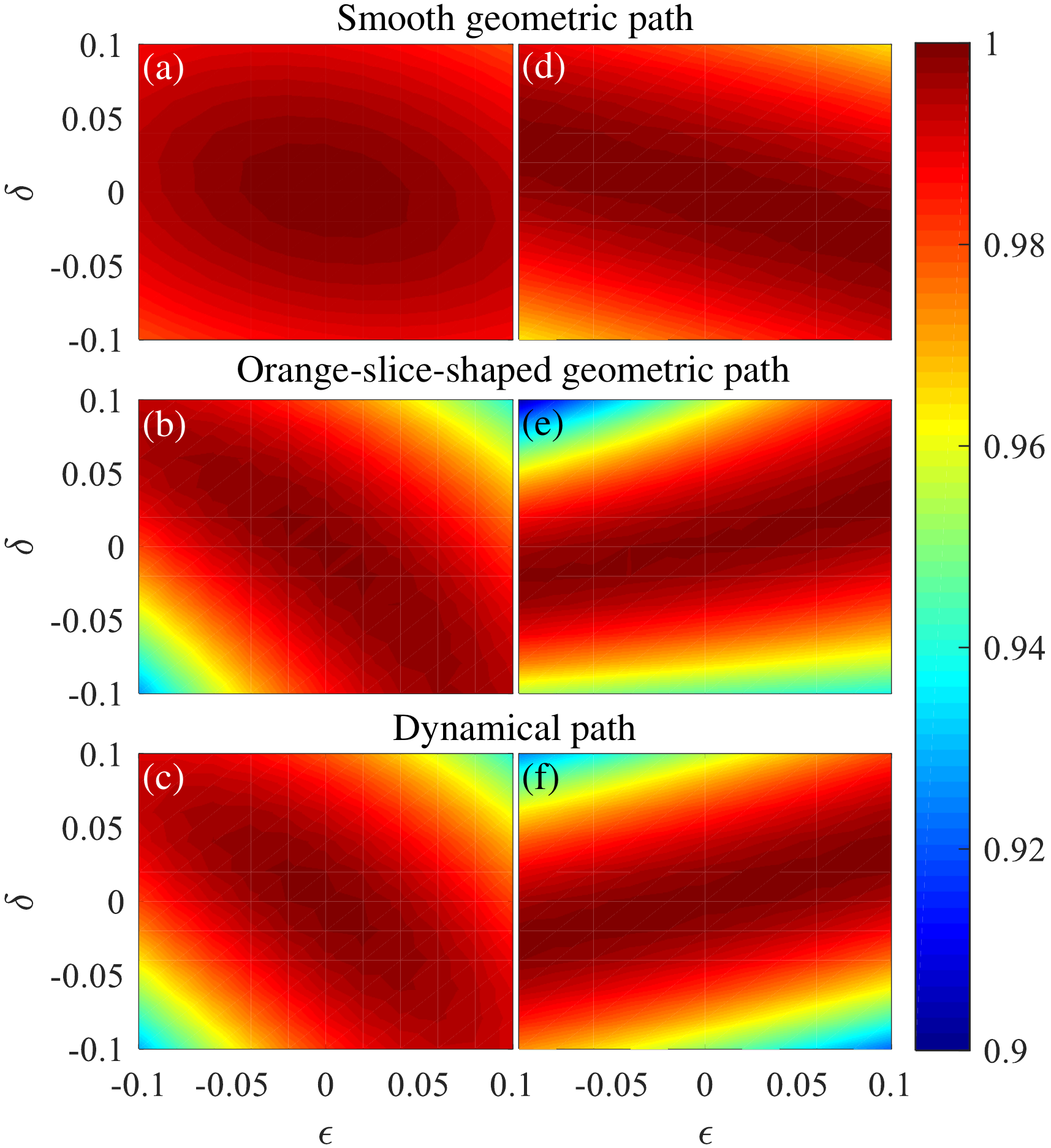}
  \caption{Gate fidelities as functions of qubit-frequency drift $\delta$ and the deviation $\epsilon$ of driving amplitude for Hadamard (-like) (left column) and $\pi/8$ (right column) gates for different schemes,   under decoherence rate of $\kappa=4\times10^{-4}\Omega_\textrm{m}$.} \label{fig3}
\end{figure}

Finally, we evaluate our gate robustness. The systematic errors, caused by qubit-frequency drift $\delta$ and the deviation $\epsilon$ of driving amplitude, and the decoherence caused by environment are the two main factors for gate infidelity. Therefore,  the system affected by systematic errors is described by
\begin{equation}\label{14}
\mathcal{H}'(t)=\frac{1}{2}\left(
  \begin{array}{cc}
    -(\Delta+\delta\Omega_{\rm m})& (1+\epsilon)\Omega(t)e^{-i\phi(t)} \\
    (1+\epsilon)\Omega(t)e^{i\phi(t)} & \Delta+\delta\Omega_{\rm m} \\
  \end{array}
\right).
\end{equation}
When $\kappa_1=\kappa_2=\kappa=4\times10^{-4}\Omega_\textrm{m}$, Figures \ref{fig3}a and \ref{fig3}d show the gate fidelities versus $\delta$ and $\epsilon$ for Hadamard-like and $\pi/8$ gates. In the same way, to demonstrate the superiority of our scheme in terms of gate-robustness, we also show the results of gate-robustness for the cyclic OSSP scheme  in Figures \ref{fig3}b and \ref{fig3}e and the dynamical scheme in Figures \ref{fig3}c and \ref{fig3}f. By comparison, our scheme is more insensitive to both $\delta$ and $\epsilon$ errors for both Hadamard-like and $\pi/8$ gate, and thus has stronger gate robustness. Note that, similar to cyclic OSSP scheme along the longitude line, a noncyclic smooth path along the longitude line can also be used to realize NGQC, see Appendix D. However, the gate-performance there is not so good. In addition, in the above comparison, the performance of Phase gate is very similar to that of the $\pi/8$ gate, thus not present here.

\section{PHYSICAL IMPLEMENTATION}
In this section, we present a realization of our scheme on a superconducting quantum circuit, consisting of capacitively coupled transmon qubits, with experimental demonstrated techniques and finally evaluate the gate performance under realistic conditions.

\subsection{Universal single-qubit geometric gates}

We first deal with the single-qubit gates. The two lowest energy levels of a transmon qubit T$_1$,
labelled by $\{|0\rangle, |1\rangle\}$, is used as our computational subspace. However, when the driving field excites the computational subspace, it will inevitably cause the simultaneous coupling of higher energy states and result in the qubit-information leakage to $|2\rangle$ or higher energy levels. Therefore, the ``derivative removal via adiabatic gate'' (DRAG) technology\cite{DRAG049, DRAG050, DRAG051} is used here to suppress this leakage. The Hamiltonian of a transmon   drives by a microwave field with frequency $\omega_d$ and adjustable phase $\phi(t)$ is
\begin{equation}\label{15}
\begin{split}
\mathcal{H}_1(t)=&\sum_{k=1}^{+\infty}\left\{\left[k\omega_1-\frac{1}{2}k(k-1)\alpha_1\right]
|k\rangle_1\langle k|\right.\\
+&\left. \left[\frac{1}{2}\Omega_D(t)\sqrt k |k-1\rangle_1\langle k|e^{i(\omega_dt-\phi(t))}+{\rm H.c.}\right] \right\},
\end{split}
\end{equation}
where $\omega_1$ and $\alpha_1$ are the qubit frequency and anharmonicity of transmon qubit T$_{1}$, respectively; and $\Omega_D(t)\!=\!\Omega(t)\!-\![i\dot\Omega(t)\!+\!\Omega(t)\dot\phi(t)+\Delta\Omega(t)]/(2\alpha_1)$ is the corrected pulse for the original pulse $\Omega(t)$ under {\rm DRAG} technology, with the detuning being $\Delta=\omega_1-\omega_d$.

\begin{figure}
\includegraphics[width=8.5cm]{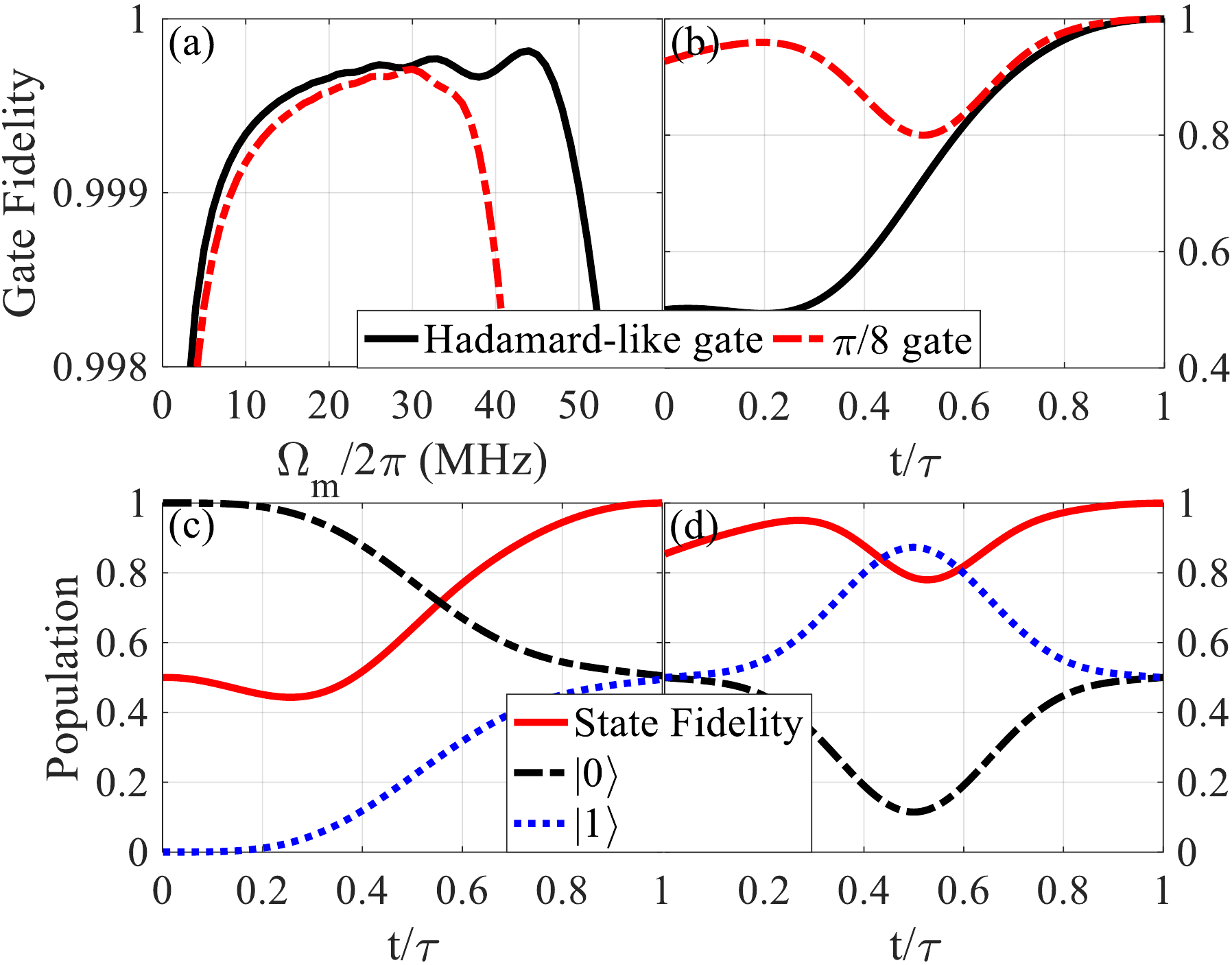}
  \caption{(a) Gate fidelities as function of $\Omega_{\rm m}$ for Hadamard-like and $\pi/8$ gates, optimal value of $\Omega_{\rm m}$ can be selected according to the highest gate fidelity. (b) Dynamics of the gate fidelity for Hadamard-like and $\pi/8$ gates using the optimal $\Omega_{\rm m}$. The qubit-state population and the state-fidelity dynamics for (c) Hadamard-like gate with initial state $|0\rangle$ and (d) $\pi/8$ gate with initial state $(|0\rangle+|1\rangle)/\sqrt2$.}\label{fig6} 
\end{figure}

Next, based on the above superconducting quantum circuit, we further simulate the performance of geometric  Hadamard-like and $\pi/8$ gates. We here also consider the qubit-state leakage to state $|2\rangle$, which is the main leakage error for transmon qubits, by choosing $k=1, 2$ in the master equation of Equation (\ref{16}) with the Hamiltonian being $\mathcal{H}_1(t)$ in Equation (\ref{15}). According to the current state-of-art experiment\cite{Exper052}, we set the qubit parameters as $\alpha_1\!=\!2\pi\times 220$ MHz and  $\kappa_1=\kappa_2=\kappa=2\pi\times\! 4$ kHz. Beside, according to Equation (\ref{13}), the detuning $\Delta/2\pi$ is set to be $-4.5$ MHz and $-1.6$ MHz for Hadamard-like and $\pi/8$ gates, respectively. Under these conditions, as shown in Figure \ref{fig6}a, there has trade-off between $\Omega_{\rm m}$  and the gate-fidelity, the optimal $\Omega_{\rm m}/2\pi$ is 44 MHz and  30 MHz for the Hadamard-like  and $\pi/8$ gates, respectively, which correspond to $\kappa/\Omega_{\rm m} \sim 10^{-4}$. The gate fidelities using the optimal $\Omega_{\rm m}$  are 99.98$\%$ and 99.97$\%$,  and the  dynamics of which is shown in Figure \ref{fig6}b. Moreover, suppose the qubit is initially in the states $\psi^1_i=|0\rangle$ and $(|0\rangle+|1\rangle)/\sqrt2$ for Hadamard-like gate and $\pi/8$ gate, the ideal final states can be $\psi^1_{\tau}=[|0\rangle\!+\!\exp(i\sqrt2\pi)|1\rangle]/\sqrt2$ and $[|0\rangle\!+\!\exp(i\pi/4)|1\rangle]/\sqrt2$, respectively. We evaluate these gates by state populations and fidelities defined by $F_S=\langle \psi^1_{\tau}|\rho_1|\psi^1_{\tau}\rangle$ with $\rho_1$ being the final simulation result of density operator. The state population and fidelity dynamics are shown in Figures \ref{fig6}c and \ref{fig6}d for Hadamard-like  and $\pi/8$ gates, respectively; and both fidelities can  reach  99.96$\%$. In addition, the performance of the geometric Phase gate is very similar to the $\pi/8$ gate, and thus is not presented here.

\subsection{Nontrivial two-qubit geometric gate}
We next proceed  to construct the nontrivial two-qubit geometric gate using two capacitively coupled transmon qubits T$_1$ and T$_2$. The Hamiltonian of two coupled qubits with coupling strength $g_{12}$ is
\begin{equation}\label{17}
\begin{split}
\mathcal{H}_2(t)=&\sum_{m=1,2}\sum_{k=1}^{+\infty}[k\omega_m-\frac{k(k-1)}{2}\alpha_m]|k\rangle_m\langle k|\\
+&g_{12}(b_{1-} {b^{\dagger}_{2-}}+{b^{\dagger}_{1-}}b_{2-}),
\end{split}
\end{equation}
where $\omega_m$ and $\alpha_m$ are the qubit frequency and anharmonicity of transmon T$_{m}$, respectively. However, due to the fixed qubit-frequency difference and the coupling strength $g_{12}$  between the two capacitively coupled transmons, the Hamiltonian in Equation (\ref{17}) has no additional degree of freedom to control its quantum dynamics, and thus leads to low gate performance. To achieve effective quantum control and better gate performance, we propose to use the experimental demonstrated parametrically tunable coupling\cite{tunable053, tunable054, tunable055} techniques. Specifically, we add an ac driving on T$_1$, which can be experimentally induced by biasing the qubit with an ac magnetic flux. This ac driving results in periodically modulating of T$_1$'s transition frequency in the form of $\omega_1(t)=\omega_1+\dot F(t)$, where $F(t)=\beta\sin[\nu t+\varphi(t)]$, with $\nu$ and $\varphi(t)$ being the frequency and phase of the longitudinal driving field, respectively.

Under this ac driving, $\mathcal{H}_{2}(t)$ in Equation (\ref{17}) can be described by replacing $\omega_1$ with $\omega_1(t)$, and we denote the new Hamiltonian as $\mathcal{H}'_{2}(t)$. Then, we move $\mathcal{H}'_{2}(t)$ to the interaction picture and use Jacobi-Anger identity, $\exp(i\beta\cos\theta)\!=\!\sum_{n}\!i^nJ_n(\beta)\!\exp(in\theta)$  with
$J_n(\beta)$ is $n$th Bessel function, to obtain the effective interaction under ac modification. Taking the first order of Bessel function terms to be the most relevant interaction, and truncate the Hamiltonian $\mathcal{H}'_{2}(t)$ into the single- and two-excitation subspaces, the reduced Hamiltonian can be written as
\begin{equation}\label{18}
\begin{split}
\mathcal{H}_{12}=&g_{12}J_1(\beta)e^{-i[\nu t+\varphi(t)]}\{|01\rangle\langle10|e^{i\Delta_1 t}\\
+&\sqrt2|02\rangle\langle11|e^{i(\Delta_1-\alpha_2)t}\\
+&\sqrt2|11\rangle\langle20|e^{i(\Delta_1+\alpha_1)t}\}+{\rm H.c.},
\end{split}
\end{equation}
where $|ml\rangle=|m\rangle_1\otimes|l\rangle_2$, and $\Delta_1=\omega_1-\omega_2$ is the qubit-frequency difference between two capacitively coupled transmons T$_1$ and T$_2$.

Setting $\Delta_L=\nu-\Delta_1$ with $|\Delta_L|\ll \{ |\nu|, |\Delta_1| \}$, one can obtain an off-resonance coupling with a slight detuning $\Delta_L$ between the states $|01\rangle$ and $|10\rangle$, while making the interactions between the  $|11\rangle$ states and the non-computational subspace, i.e., \{$|02\rangle$,$|20\rangle$\}, in a large-detuning way, and thus the leakage of the quantum information to the non-computational subspace from state $|11\rangle$ can be greatly suppressed. Specifically, considering the single-excitation subspace $\{|01\rangle, |10\rangle \}$ of $\mathcal{H}_{12}$, and rotating to the framework with operator $U_{\Delta_L}\!=\!\exp[-i\Delta_L(|01\rangle\langle01|-|10\rangle\langle10|)t/2]$, we can obtain an effective two-level type Hamiltonian
\begin{equation}\label{19}
\begin{split}
\mathcal{H}_{12}'=&\frac{1}{2}\left(
  \begin{array}{cc}
    -\Delta_L& g'_{12}e^{-i\varphi(t)} \\
    g_{12}'e^{i\varphi(t)} & \Delta_L \\
  \end{array}
\right),
\end{split}
\end{equation}
which is in the same form of the single-quit case, see Equation (\ref{01}), where $g_{12}'\!\!=\!\!2g_{12}J_1(\beta)$ is the effective coupling strength that is  tunable via $\beta$.

\begin{figure}
\includegraphics[width=8.5cm]{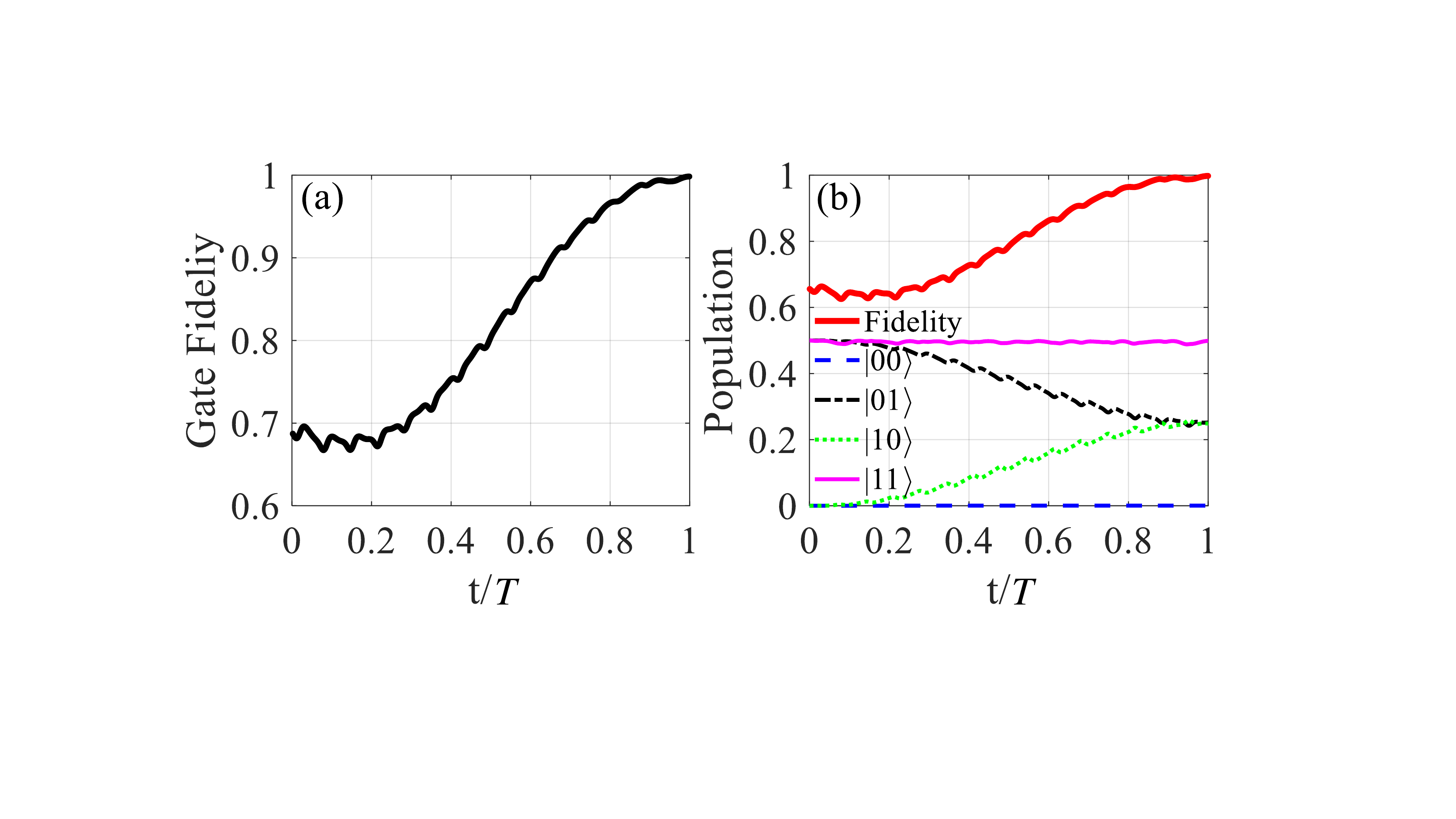}
\caption{The performance of $\sqrt{{\rm iSWAP}}$-like gate. (a) Dynamics of the gate fidelity. (b) Qubit-state population and state-fidelity dynamics, with the initial state $(|01\rangle+|11\rangle)/\sqrt2$.}\label{fig7}
\end{figure}

Next, we proceed to construct the nontrivial two-qubit geometric $\sqrt{{\rm iSWAP}}$-like gate in the subspace $\mathcal{P}\!=\!\{|00\rangle,\!|01\rangle,\!|10\rangle,\!|11\rangle\}$. In the single-excitation subspace $\{|01\rangle$, $|10\rangle\}$, similar to the single-qubit case, by setting $\Gamma=\pi/2, \xi(0)=-\pi/4$ and $\chi=\pi/4$ in Equation (\ref{12}), we can also obtain {\small $R_z(\xi'_{-})\left(\begin{array}{cc}1 & i \\i & 1 \\\end{array}\right)/\sqrt2$} with $\xi'_{-}=(\sqrt2-1)\pi$. Meanwhile, in the subspace $\{|00\rangle, |11\rangle\}$, we can get an identity operator under the premise that no leakage of state $|11\rangle$. Therefore, a two-qubit $\sqrt{{\rm iSWAP}}$-like geometric gate can be achieved.

Finally, we turn to test the performance of this two-qubit gate. To simulate the decoherence effects of our considered two coupled transmons, we also use the master equation method with
\begin{equation}
\begin{split}
\dot\rho_2=&-i[\mathcal{H}'_{2}(t),\rho_2]+\kappa_1\mathcal{A}(b_{1-}) +\kappa_2\mathcal{A}(b_{1z})\\
&+\kappa'_1\mathcal{A}(b_{2-})+\kappa'_2\mathcal{A}(b_{2z}),
\end{split}
\end{equation}
where $\kappa'_1$ and $\kappa'_2$ are the decay and dephasing rates of the second transmon, respectively. For a general initial state $|\psi^2_i\rangle\!=\!(\cos\theta_1|0\rangle_1\!+\!\sin\theta_1|1\rangle_1)\otimes (\cos\theta_2|0\rangle_2\!+\!\sin\theta_2|1\rangle_2)$, the ideal final state is $|\psi^2_T\rangle=\sqrt{{\rm iSWAP}}$-like$|\psi^2_i\rangle$. Here, we define $F^2_G=\int_0^{2\pi}\int_0^{2\pi}\langle\psi^2_T|\rho_2|\psi^2_T\rangle d\theta_1d\theta_2/(4\pi^2)$\cite{fidelity} as the formula of two-qubit gate-fidelity and the integration is numerically performed for 10001 initial states with $\theta_1$ and $\theta_2$ evenly distributed over $[0, 2\pi]$. Besides, we set the coupling strength as $g_{12}=2\pi\times8$ MHz, $\alpha_2=2\pi\times 180$ MHz, decay and dephasing rates as $\kappa=\kappa'_{1}=\kappa'_{2}=2\pi\times 4$ kHz. In addition, to reduce the influence from the high-order oscillating terms, optimizing the driving parameter $\beta$ for high gate-fidelity is necessary. Considering the current technology, we choose the optimized $\beta=1.3$ and  $\Delta_1=2\pi\times345$ MHz. Under these settings, we can get $\Delta_L\approx-2\pi\times8.4$ MHz. Accordingly, the gate-fidelity can reach to 99.84$\%$ as shown in Figure \ref{fig7}a. In Figure \ref{fig7}b, we show the qubit-state population and state-fidelity dynamics with the initial state being $(|01\rangle+|11\rangle)/\sqrt2$. We can see that the oscillation of state $|11\rangle$ is very slight, which means the leakages from the state $|11\rangle$ is indeed be effectively suppressed. At the final time, the state fidelity can reach to 99.78$\%$.

\section{conclusion}
In summary, we proposed a scheme for robust and high-fidelity geometric quantum gates, using the noncyclic smooth geometric paths. Our path-design scheme has the shorter evolution time and is also experimental friendly, i.e., do not require parameter mutation and time-dependent modulation of the detuning. Therefore,  the gate performance in our scheme can be better than conventional dynamical gates, which thus is a promising alternative towards scalable and fault-tolerant quantum computation.

\appendix
\section*{Appendix}
\section{Calculation of the geometric phase}
In this Appendix, we present the detail of calculating the noncyclic  geometric phases, following Ref.\cite{PP049}. Here, we take $|\psi_{1}(t)\rangle$ as an example, the total relative phase from the initial state $|\psi_1(0)\rangle$ to final state $U(\tau)|\psi_1(0)\rangle$ is
\begin{equation}
\begin{split}
\label{07}
\gamma_t(\tau)=&{\rm arg}\langle \psi_{1}(0)|U(\tau)|\psi_{1}(0)\rangle\\
=&{\rm arg}\langle \psi_{1}(0)|e^{i\gamma(\tau)}|\psi_{1}(\tau)\rangle\\
=&\gamma(\tau)+{\rm arg}\langle \psi_{1}(0)|\psi_{1}(\tau)\rangle.
\end{split}
\end{equation}
Removing the dynamical phase in Equation (\ref{08}) from the total phase, the remaining is the Pancharatnam geometric phase\cite{SB034, NNGQC039, PP049}
\begin{eqnarray} \label{09}
\gamma_g(\tau)&=& \gamma_t(\tau)-\gamma_d(\tau) \notag\\
&=& \gamma(\tau)-\gamma_d(\tau)+{\rm arg}\langle \psi_{1}(0)|\psi_{1}(\tau)\rangle\\
&=& i\int_0^\tau \langle \psi_{1}(t)|\frac{\partial}{\partial t}|\psi_{1}(t)\rangle dt
+{\rm arg}\langle \psi_{1}(0)|\psi_{1}(\tau)\rangle, \notag
\end{eqnarray}
which is gauge invariant. Specially, for cyclic path,  $|\psi_{1}(0)\rangle=|\psi_{1}(\tau)\rangle$, i.e., ${\rm arg}\langle \psi_{1}(0)|\psi_{1}(\tau)\rangle=0$, and thus  Equation (\ref{09}) reduces to the conventional cyclic Aharonov-Anandan phase\cite{AAP013}. Therefore, the phase accumulated through noncyclic path in this work is a generalization of the cyclic case. In addition, here an auxiliary geodesic line closing the initial and final states is introduced to explain the geometric property\cite{SB034, PP049}. Then, the geometric phase of noncyclic path $\mathcal{C}_1$ with it's geodesic $\mathcal{C}_2$ can be calculated, see Equation (\ref{11}).

\section{Geometric phases from cyclic path}
In previous works of NGQC, cyclic evolution is often used when constructing a set of universal geometric quantum gates. Taking the resonant coupling scheme $(\Delta=0)$ for example, to eliminate the dynamical phase, the path is choose to evolve along the longitude line so that $\dot\xi=0$, but at the same time, the geometric phase is also null. In this case, two longitude lines, corresponding to different azimuth angle $\xi_0$ and $\xi_1$, are used to form a close path with a solid angle. And the geometric phase can be induced on the mutation point (for example, the south pole). Thus, the OSSP geometric evolution has been designed,
where the trajectory is divided into three segments, as shown in Figure \ref{fig1}a (Path 2). The evolution operator is
\begin{equation}\label{C1}
\begin{split}
&U_c(\gamma_g, \chi_0, \xi_1)\\
 &=\left(
  \begin{array}{cc}
  \cos\gamma_g+i\sin\gamma_g\cos\chi_0&i\sin\gamma_g\sin\chi_0 e^{-i\xi_1}\\
   i\sin\gamma_g\sin\chi_0 e^{i\xi_1}&\cos\gamma_g-i\sin\gamma_g\cos\chi_0\\
  \end{array}\right),\\
\end{split}
\end{equation}
where $\gamma_g=\xi_1-\xi_0$ is the total geometric phase produced when the path mutates at the south pole, and the dynamical phase in this process is always zero. Equation (\ref{C1}) is one of the design methods of Equation (\ref{06}), i.e., $\Gamma=\gamma_g$, and universal single-qubit gates can be realized\cite{NG019,NG020}. In this construction, the parameters of the Hamiltonian satisfying
\begin{equation}
\centerline{$\left\{
        \begin{array}{lll}
           t\in[0,\tau_1):&\phi=\xi_0-\frac{\pi}{2},&\int_0^{\tau_1}\Omega(t)dt=\chi_0, \\
           t\in[\tau_1,\tau_2]:&\phi=\xi_1+\frac{\pi}{2},&\int_{\tau_1}^{\tau_2}\Omega(t)dt=\pi,\\
           t\in(\tau_2,\tau]:&\phi=\xi_0-\frac{\pi}{2},&\int_{\tau_2}^{\tau}\Omega(t)dt=\pi-\chi_0.
        \end{array}
    \right.$}\\
\end{equation}
However, in this case, the qubit parameters need to be mutated, which may increase the control complexity and thus cause control error in an experimental practice.

\section{Dynamical gates}
Previously, compared with dynamical quantum gates, the pursuit of better performance of geometric gates is usually limited by the complex multiple levels/qubits interactions and the required longer gate duration. Therefore, we here take dynamical gates as typical reference to highlight higher fidelity and stronger robustness of our geometric gates. General dynamical evolution operator constructed by Hamiltonian in Equation (\ref{01}) with $\Delta=0$ and $\phi=\phi_d$ is a constant can be written as
\begin{equation}\label{D1}
\begin{split}
&U_d(\theta_d, \phi_d)=e^{-i\int_0^\tau\mathcal{H}(t)dt}\\
 &=\left(
  \begin{array}{cc}
  \cos(\theta_d/2)&-i\sin(\theta_d/2)e^{-i\phi_d}\\
   -i\sin(\theta_d/2)e^{i\phi_d}&\cos(\theta_d/2)\\
  \end{array}\right),
\end{split}
\end{equation}
where parameters $\theta_d\!=\!\int_0^\tau\Omega(t)dt$, and $\phi_d$ is a constant that ensures the geometric phase is zero. In this way, the arbitrary dynamical X-, Y- and Z-axis rotation operations can all be realized by $R^d_x(\theta_x)=U_d(\theta_x, 0)$, $R^d_y(\theta_y)=U_d(\theta_y, \pi/2)$ and $R^d_z(\theta_z)=U_d(\pi/2, \pi)U_d(\theta_z,-\pi/2)U_d(\pi/2,0)$, respectively. In addition, a set of universal single-qubit gates, i.e., Hadamard, Phase and $\pi/8$ gates can all be realized as $U_d(\pi, \pi)U_d(\pi/2, \pi/2)$, $R^d_z(\pi/2)$ and $R^d_z(\pi/4)$, respectively.

\begin{figure}
\includegraphics[width=8cm]{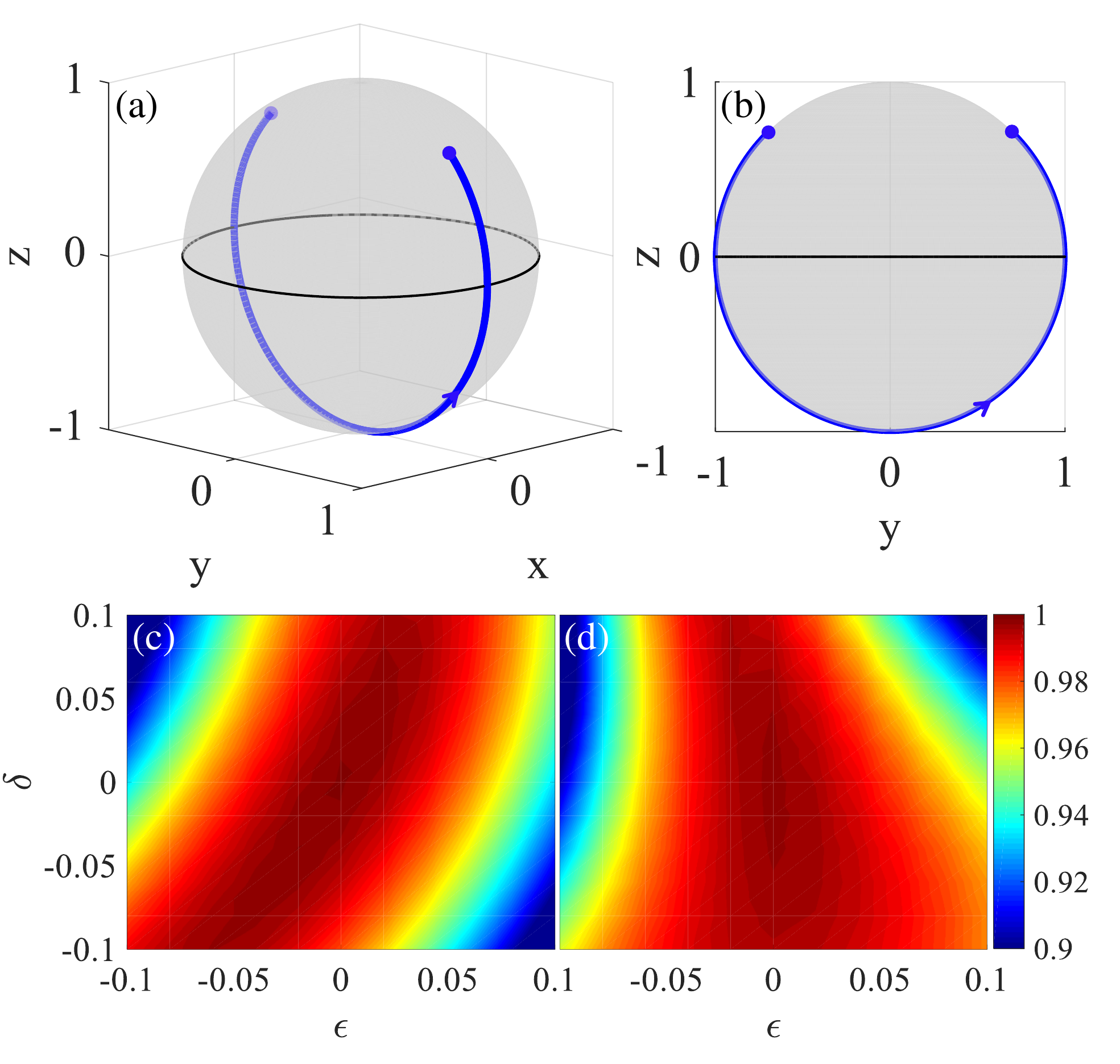}
\caption{(a) The trajectory of smooth geometric path along the longitude line. (b) The top view in z-y plane of the the trajectory in (a). Gate fidelity as functions of qubit frequency drift $\delta$ and the deviation $\epsilon$ of driving amplitude for (c) Hadamard-like gate and (d) $\pi/8$ gate under the unified decoherence rate $\kappa=4\times 10^{-4} \Omega_\textrm{m}$.} \label{fig8}
\end{figure}

\section{Noncyclic geometric path along longitude}
From section II, we conclude that our noncyclic smooth geometric path along the latitude line shows stronger resistance to the systematic errors and decoherence effect. Then a problem arise, similar to the OSSP scheme, how the gate performance will be if the noncyclic path goes along longitude line? So, we also design such a scheme, as illustrated in Figures \ref{fig8}a and \ref{fig8}b. In this case, the trajectory needs to go through the south pole ($\chi\!\!=\!\!\pi$) where a $-\pi$ geometric phase is induced while the dynamical phase still equals zero. Except the south pole, as $\dot\xi=0$ and $\Delta=0$, the dynamical phase $\gamma_d=0$. Therefore, based on this  geometric  path, we can construct a universal single-qubit gates by setting $\Gamma=-\pi/2$, $\chi(0)\!=\!\chi(\tau)\!=\!\chi_0$, $\xi_{-}\!=\!\pi$, $\xi_{+}\!=\!2\xi_0\!+\!\pi$ in Equation (\ref{06}). And,  the corresponding evolution operator is
\begin{equation}\label{017}
\begin{split}
U_n(\chi_0, \xi_0)=
 &\left(
  \begin{array}{cc}
  \cos\chi_0&\sin\chi_0e^{-i\xi_0}\\
  -\sin\chi_0e^{i\xi_0}&\cos\chi_0\\
  \end{array}\right).\\
\end{split}
\end{equation}
In this way, arbitrary rotation operations can be realized by $R^n_x(\theta_x)=U_n(\theta_x/2, \pi/2)$, $\!R^n_y(\theta_y)=U_n(\theta_y/2, \pi)$ and $R^n_z(\theta_z)=U_n(\pi/4, -\pi/2)U_n(\theta_z/2, 0)U_n(\pi/4, \pi/2)$. In addition, an universal set of quantum gates, i.e., the Hadamard, Phase and $\pi/8$ gates, can be achieved by $U_n(\pi/2, \pi/2)U_n(\pi/4, \pi)$, $R^n_z(\pi/2)$ and $R^n_z(\pi/4)$ with the effective pulse areas are $S=5\pi/4$, $2\pi$ and $19\pi/8$, respectively. Figures \ref{fig8}c and \ref{fig8}d show  the gate fidelities versus qubit-frequency drift $\delta$ and the deviation $\epsilon$ of driving amplitude for the Hadamard and $\pi/8$ gates, under the unified decoherence rate of $\kappa=4\times10^{-4}\Omega_\textrm{m}$. Overall, the noise-resilience of the gates based on the path along  the longitude line is not so good as that of based on the latitude line in the maintext.

\section*{Acknowledgments}

This work is supported by the Development Program of GuangDong Province (No. 2018B030326001), the National Natural Science Foundation of China (No. 11874156), the National Key R$\&$D Program of China (No. 2016YFA0301803), and Science and Technology Program of Guangzhou (No. 2019050001).

\end{document}